# Time Series Classification for Locating Forced Oscillation Sources

Yao Meng, *Student Member, IEEE*, Zhe Yu, *Member, IEEE*, Ning Lu, *Senior Member*, *IEEE*, Di Shi, *Member*, *IEEE*

*Abstract*— **Forced oscillations are caused by sustained cyclic disturbances. This paper presents a machine learning (ML) based time-series classification method that uses the synchrophasor measurements to locate the sources of forced oscillations for fast disturbance removal. Sequential feature selection is used to identify the most informative measurements of each power plant so that multivariate time series (MTS) can be constructed. By training the Mahalanobis matrix, we measure and compare the distance between the MTSs. Templates for representing each class is constructed to reduce the size of training datasets and improve the online matching efficiency. Dynamic time warping (DTW) algorithm is used to align the out-of-sync MTSs to account for oscillation detection errors. The algorithm is validated on two test systems: the IEEE 39-bus system and the WECC 179-bus system. When a forced oscillation occurs, MTSs will be constructed by designated PMU measurements. Then, the MTSs will be classified by the trained classifiers, the class membership of which corresponds to the location of each oscillation source. Simulation results show that the proposed method can be used online to identify the forced oscillation sources with high accuracy. The robustness of the proposed algorithm in the presence of oscillation detection errors is also quantified.**

## I. INTRODUCTION

LOW frequency oscillations can compromise the power quality, cause damage to system equipment, and lead to catastrophic backouts [1]. In general, the oscillations can be classified into two categories: poorly damped natural oscillations (PDNO) and forced oscillations (FO).

PDNOs are normally caused by short-lived disturbances in systems with weak or negative damping. Examples of such disturbances can originate from power variations on long transmission lines or control actions from high-gain fast exciters or control devices with poorly tuned control settings [2],[3]. PDNOs can be suppressed by tuning parameters of power system stabilizers (PSSs) or adjusting power flows on intertie lines [4]. Bonneville Power Administration (BPA) has established Dispatcher Standing Order 303 to reduce the amount of the tie-line flow when the system inertia drops to a dangerously low level.

FOs are normally casued by resonances excited by external periodic disturbances. An FO can happen in a well-damped system at a frequency close or equal to the intrinsic system frequency [5]. Because periodic small disturbances can be amplified by resonance, as long as the periodic driving source exists, the oscillation will persist and may propagate rapidly to the whole system. A FO source can be a malfunctional PSS, a generator turbine with mechanical oscillations, a load with cyclic changes [6], [7]. An emerging FO source is wind turbines. Periodical power fluctuations caused by wind shear and tower shadow effects or vibrations of floating offshore wind turbines [8], [9] have been observed to cause FO. Thus, it is expected that the occurrence of FOs in high-wind-penetration grids will increase significantly. Because FOs are caused by resonance, the oscillation will persist as long as the FO source still presents. Therefore, locating and removing the FO source is crucial to quench an FO.

After the wide deployment of Phasor Measurement Units (PMUs), high-resolution synchronized measurements are used to detect, diagnose, and locate oscillations previously invisible to the conventional supervisory control and data acquisition (SCADA) systems [10]. As introduced in [11], researchers have reviewed a few forced oscillation source location methods using PMU measurements. In [12], Markham and Liu introduced a traveling wave based method for integrating PMU measurements at different places and identifying the generator with the earliest presence of oscillation as a source. The authors assume that the electric wave speed throughout the network is the same. This assumption may not hold in reality because electric wave speed varies between 50% to 80% speed of the light depending on the network parameters. In [13],[14], the energy flow direction is calculated based on the transient energy function to locate the forced oscillation source in over 50 practical events. This energy-based method tracks the flow of effective transient energy, thus being model independent. However, it is hard to distinguish between a true source bus and a bus having a negative damping contribution. As illustrated in [15], the generator caused FO can be characterized as an effective current source in addition to its effective admittance. Prior knowledge of generator model structures is needed to implement this method.

This work is funded by SGCC Science and Technology Program under project "AI based oscillation detection and control" with contract number SGJS0000DKJS1801231.

Yao Meng was with the Electrical & Computer Engineering Department, North Carolina State University, Raleigh, NC 27606, USA and GEIRI North America, San Jose, CA 95134, USA. He is now with China Electric Power Planning & Engineering Institute, Beijing, China. (email: ymeng@eppei.com).

Zhe Yu, Di Shi are with GEIRI North America, San Jose, CA 95134, USA. (emails: zhe.yu@geirina.net, di.shi@geirina.net).

Ning Lu is with the Electrical & Computer Engineering Department and Future Renewable Energy Delivery and Management (FREEDM) Systems Center, North Carolina State University, Raleigh, NC 27606 USA. (email: nlu2@ncsu.edu)

There are two issues in the existing methods. First, the data used for FO source detection were selected from the very beginning phase of an FO with an underlying assuming that an FO can be accurately detected after it occurs, which is not a valid assumption in real-time FO detection. Second, most authors discussed only the accuracy of their algorithms without considerations made to meet the computational time requirements. Therefore, in this paper, we propose a machine learning based method for locating FO sources in real-time considering both the accuracy and computational time requirements.

We formulate the problem as a multivariate time series (MTS) classification problem with each class membership corresponding to the location of the FO source. The most informative PMU measurements for constructing MTS will be identified using Sequential Feature Selection. Mahalanobis metrics are used to represent the distance between the MTSs. To meet real-time identification needs, templates representing each class are constructed in offline training in order to improve the algorithm efficiency. Dynamic time warping (DTW) is applied so that MTS of different phases can be appropriately compared, relaxing the need to use data in the exact beginning phase of an oscillation.

The rest of the paper is structured as follows. Section II introduces the offline training process for MTS classification. Section III describes DTW and the improved $k$-nearest-neighbors ($k$-NN) method for resolving the MTS out-of-sync issue. Section IV presents the simulation results and performance evaluation. Conclusions and future work are summarized in Section V.

## II. MTS CLASSIFICATION

In this section, we will introduce the mechanism of the forced oscillation and the MTS classification process.

### A. Mechanisms of the Forced Oscillation

As introduced in [5], forced oscillations are caused by periodically disturbances at the system resonance frequency. In a single-machine infinite-bus (SMIB) system, a periodical external disturbance injected to the mechanical power of a generator, $\Delta P_{FO}$, can be represented as

$$\Delta P_{FO} = \Delta P_{FO}^{Max} \sin(\omega_{FO} t) \qquad (1)$$

where $\Delta P_{FO}^{Max}$ is the amplitude and $\omega_{FO}$ is the angular frequency. The resultant change in generator rotor speed can be calculated from

$$\frac{T_J}{\omega_0}\frac{d^2\Delta\delta}{dt^2} + \frac{K_D}{\omega_0}\frac{d\Delta\delta}{dt} + K_s\Delta\delta = \Delta P_{FO} - \Delta P_e \qquad (2)$$

where $T_J$ is the generator inertia time constant, $\omega_0$ is the system reference angular frequency with $f_0 = 60$ Hz , $\delta$ is the generator power angle, $K_D$ is the generator damping factor, $K_s$ is the generator synchronous torque coefficient, $\Delta P_e$ is generator electromagnetic power output, and $\Delta P_T$ is the generator mechanical power output.

If only the generator mechanical power is perturbed (*i.e.*, $\Delta P_e = 0$), (2) becomes

$$\Delta\ddot{\delta} + 2\xi\omega_n\Delta\dot{\delta} + \omega_n^2\Delta\delta = \frac{\omega_0}{T_J}\Delta P_{FO}^{Max}\sin(\omega t) \qquad (3)$$

$$\omega_n = \sqrt{\omega_0 K_s / T_J} \qquad (4)$$

$$\xi = \frac{K_D}{(2\sqrt{\omega_0 K_s T_J})} \qquad (5)$$

where $\omega_n$ is the system natural frequency and $\xi$ is the damping ratio. The solution of (3) consists of a general solution and a particular solution. The particular solution is

$$\Delta\delta(t) = B_T\sin(\omega t - \phi) \qquad (6)$$

where

$$B_T = \frac{\frac{\omega_0}{T_J}\Delta P_{FO}^{Max}/K_s}{\sqrt{[1-(\omega_{FO}/\omega_n)^2]^2+(2\xi\omega_{FO}/\omega_n)^2}} \qquad (7)$$

From (7), we can see that when $\omega_{FO} = \omega_n$, the periodical driving force $\Delta P_{FO}$ will bear the same frequency as the generator natural frequency, resulting a resonance with a maximum $B_T$ . When this power oscillation propagates to the rest of the grid, an FO will occur. In an FO event, PMUs located at other buses will bear a distinct sequential time series relationship, which is correlated to the distance to the FO source.

### B. MTS Classification

There are two main steps in MTS classification for locating the FO source: offline training and online identification.

To prepare the training data, we made the following assumptions:

1. On each generator bus, there is one and only one PMU installed. Thus, there are $N$ PMUs installed for an $N$-generator system.
2. Consider only cases with a single FO source. Cases with multi-FO sources are ignored because the occurrence of such events is rare.
3. An FO source will be modeled as disturbances to the mechanical power of a generator. This is because the periodic disturbance is often on a generator [13].

Because of assumptions 1-3, the number of the FO source location is equal to the number of generator buses.

As the first step, $J$ typical system operation conditions are generated to emulate practical situations, *i.e.* load fluctuations. For a given operation condition $j$, we perturb one generator bus at a time until all $N$ generators are perturbed. Thus, when the $i^{th}$ generator bus is disturbed, the corresponding electromechanical transients recorded by all of the $N$ PMU located at each generator bus will form a MTS matrix, $\boldsymbol{X}$, which is represented as

$$\mathbf{X}_i^j(h,p) = \begin{bmatrix} x_1(1) & x_2(1) & \cdots & x_p(1) \\ x_1(2) & x_2(2) & \cdots & x_p(2) \\ \vdots & \vdots & \ddots & \vdots \\ x_1(h) & x_2(h) & \cdots & x_p(h) \end{bmatrix}_{H\times P} \qquad (8)$$

where $h$ is the index of the number of sampling points, $p$ is index of the PMU channels, $j$ represents the different system operation conditions, and $i$ denotes the location of generator being perturbed, *i.e.* oscillation source. Each column in $X$ represents a time series data of a selected PMU channel and each row represents a snapshot of all selected PMU channels. Because we assume that the oscillation source is located on a generator bus, $i$ is indexed from 1 to $N$. To locate the oscillation source, we consider the MTS with the same subscript values to be in the same class.

Because each PMU records several data channels (e.g. power angle, $\delta_i$, real and reactive power, $P_i$ and $Q_i$), a feature selection process is needed to determine the $T$ most informative channels in order to construct the MTS. Thus, $P = N \times T$ and $H$ is the total number of samples in the time series of the selected PMU channel.

### 1) Mahalanobis Distance Calculation

Mahalanobis distance is a standard measure of the distance between two points in the multivariate space [16]. In this paper, we use the Mahalanobis distance to characterize the distance between two time series data sets. Let $x$ and $y$ be the $h^{\text{th}}$ row of two different MTS matrices $\mathbf{X}_{i_1}^{j_1}$ and $\mathbf{X}_{i_2}^{j_2}$:

$$x = \mathbf{X}_{i_1}^{j_1}(h, 1:P) \tag{9}$$

$$y = \mathbf{X}_{i_2}^{j_2}(h, 1:P) \tag{10}$$

Note that $x$ and $y$ represents the $h^{\text{th}}$ data point of all $P$ selected PMU channels when perturbing Generator $i_1$ and $i_2$ under operation condition $j_1$ and $j_2$, respectively.

Then, the Mahalanobis distance between $x$ and $y$ can be calculated as:

$$d_M(x, y) = (x - y)^T \mathbf{M}(x - y) \tag{11}$$

$$\mathbf{M} = \mathbf{U}\Sigma\mathbf{U}^T \tag{12}$$

where $\mathbf{M}$ is the Mahalanobis matrix, a symmetry Positive Semi-Definite (PSD) matrix. The dimension of $\mathbf{M}$ is $N \times N$. If $\mathbf{M} = \mathbf{I}$, Mahalanobis distance degenerates to the standard Euclidean distance. $\mathbf{U}$ is a unitary matrix that satisfies $\mathbf{U}\mathbf{U}^T = \mathbf{I}$. $\Sigma$ is a diagonal matrix with singular values on the diagonal.

Thus, (10) can also be written as

$$d_M(x, y) = (x - y)^T \mathbf{U}\Sigma\mathbf{U}^T(x - y)$$
$$= (\mathbf{U}^T x - \mathbf{U}^T y)^T \Sigma(\mathbf{U}^T x - \mathbf{U}^T y) \tag{13}$$

where $\mathbf{U}$ is used to remove the correlation between variates so that the original space can be mapped into a new coordinate system and $\Sigma$ servers as the weighting for the new variates.

The distance between $\mathbf{X}_{i_1}^{j_1}$ and $\mathbf{X}_{i_2}^{j_2}$ is defined as

$$D_M\left(\mathbf{X}_{i_1}^{j_1}, \mathbf{X}_{i_2}^{j_2}\right) = \sum_{h=1}^{H} d_M\left(\mathbf{X}_{i_1}^{j_1}(h, 1:P), \mathbf{X}_{i_2}^{j_2}(h, 1:P)\right) \tag{14}$$

where $\mathbf{X}_{i_1}^{j_1}(h)$ and $\mathbf{X}_{i_2}^{j_2}(h)$ denotes the $h^{\text{th}}$ row in $\mathbf{X}_{i_1}^{j_1}$ and $\mathbf{X}_{i_2}^{j_2}$, respectively.

### C. Optimal Mahalanobis Matrix Selection

Compared with traditional distance measuring methods, Mahalanobis distance has the advantages of eliminating the need to standardize the data [17]. By searching for an optimal $\mathbf{M}$ that shortens the distance between the MTSs in the same class, the accuracy of member identification for locating the FO source can be significantly improved.

Our main idea is to use metric learning on the labeled training data to find an $\mathbf{M}$ that amplifies the influence of the location of the oscillation source while minimizing the influence of different system operation conditions. For example, for any two given sets of time series data in the MTS matrix, $\mathbf{X}_{i_1}^{j_1}$ and $\mathbf{X}_{i_2}^{j_2}$, the $\mathbf{M}$ learned should shorten the distance if $i_1 = i_2$, and lengthen the distance if $i_1 \neq i_2$. There are three constraints in metric learning: class label, pairwise label, and triplet label. As discussed and proved in [18-21], the triplet label has the best performance with the weakest constraints, so the triplet label is selected for metric learning.

First, we select a triplet label $\{\mathbf{X}_{i_1}^{j_1}, \mathbf{X}_{i_2}^{j_2}, \mathbf{X}_{i_3}^{j_3}\}$, in which $\mathbf{X}_{i_1}^{j_1}$ and $\mathbf{X}_{i_2}^{j_2}$ are in the same class (i.e., $i_1 = i_2, j_1 \neq j_2$) and $\mathbf{X}_{i_3}^{j_3}$ is in a different class (i.e. $i_3 \neq i_1$). Then, we define the constraint. Because the goal of metric learning is to find $\mathbf{M}_{opt}$ that shortens the distance between instances in the same class (i.e., $D_M\left(\mathbf{X}_{i_1}^{j_1}, \mathbf{X}_{i_2}^{j_2}\right)$) and lengthens the distances between instances from the different classes ($D_M\left(\mathbf{X}_{i_1}^{j_1}, \mathbf{X}_{3}^{j_3}\right)$ and $D_M\left(\mathbf{X}_{i_2}^{j_2}, \mathbf{X}_{i_3}^{j_3}\right)$), the following constraint needs to be satisfied for all triplet constraints $\{\mathbf{X}_{i_1}^{j_1}, \mathbf{X}_{i_2}^{j_2}, \mathbf{X}_{i_3}^{j_3}\}$:

$$\min\left(D_M\left(\mathbf{X}_{i_1}^{j_1}, \mathbf{X}_{3}^{j_3}\right), D_M\left(\mathbf{X}_{i_2}^{j_2}, \mathbf{X}_{i_3}^{j_3}\right)\right) - D_M\left(\mathbf{X}_{i_1}^{j_1}, \mathbf{X}_{i_2}^{j_2}\right) > \rho \tag{15}$$

where $\rho \geq 0$ represents the targeted margin.

In the training phase, we need to select as many triplet constraints as possible in order to escape local minimums. However, because the number of triplets is the cubic of the number of the training samples, it is not practical to use all possible triplet combinations. Thus, a dynamic triplets building strategy based on the current learned $\mathbf{M}$ is proposed by Mei et al. in [22]. This method allows us to choose the most useful triplets at the boundaries of different classes instead of random selection of triplets. For example, we can choose $\mathbf{X}_{i_2}^{j_2}$ to be the data point that has the largest distance in the same class of $\mathbf{X}_{i_1}^{j_1}$, while $\mathbf{X}_{i_3}^{j_3}$ as the one with the nearest distance to $\mathbf{X}_{i_1}^{j_1}$ in a different class.

The iterative process for updating $\mathbf{M}$ is explained as follows. First, select a triplet constraint as the input to the learning process. If (15) is satisfied, go to next iteration. Otherwise, update $\mathbf{M}$ to reduce the loss function $L(\mathbf{M})$, which is defined as

$$L(\mathbf{M}) = \rho + D_M\left(\mathbf{X}_{i_1}^{j_1}, \mathbf{X}_{i_2}^{j_2}\right) - D_M\left(\mathbf{X}_{i_1}^{j_1}, \mathbf{X}_{i_3}^{j_3}\right) \tag{16}$$

At each iteration, $\mathbf{M}$ is updated by minimizing the loss function. To guarantee the stability of the learning process, a regularization term that restricts the divergence of matrices is added using the LogDet matrix divergence [23] calculated as

$$\text{Div}(\mathbf{M}_t, \mathbf{M}_{t+1}) = \text{tr}(\mathbf{M}_t \mathbf{M}_{t+1}^{-1}) - \log(\det(\mathbf{M}_t \mathbf{M}_{t+1}^{-1})) - N \tag{17}$$

where the function "tr()" stands for the trace of a matrix and $N$ is the dimension of $\mathbf{M}$, $\mathbf{M}_t$ is the current learned $\mathbf{M}$ matrix.

Thus, the $\mathbf{M}$ updating process [22] can be expressed as:

$$\mathbf{M}_{t+1} := \underset{M \succ 0}{\arg\min} \{Div(\mathbf{M}_t, \mathbf{M}) + \eta L(\mathbf{M})\} \tag{18}$$

where $\eta$ is the parameter balancing the loss function $l(\mathbf{M})$ and the regularization term $Div(\mathbf{M}_t, \mathbf{M})$. The $\mathbf{M}_{t+1}$ after the final iteration is called $\mathbf{M}_{opt}$.

If $\eta$ is too large, the learning process minimizes distances between the same class and maximizes the distance between different classes for each triplet. This can result in an unstable updating. If $\eta$ is too small, the divergence between each iteration will be limited, leading to a slow learning process.

The objective function reaches its minimum when the gradient is zero. By setting the gradient of function $Div(\mathbf{M}_t, \mathbf{M}) + \eta l(\mathbf{M})$ as zero, we have

$$M_{t+1} = (M_t^{-1} + \eta(P_t P_t^T - Q_t Q_t^T))^{-1} \quad (19)$$

$$P_t = X_{i_1}^{j_1} - X_{i_2}^{j_2} \quad (20)$$

$$Q_t = X_{i_1}^{j_1} - X_{i_3}^{j_3} \quad (21)$$

where $P_t$ denotes the difference between instances in the same class and $Q_t$ denotes the difference between instances in the different class. To reduce the computation burden of matrix inverse, the standard Woodbury matrix identity introduced in [22] is used to solve (17) as follows:

$$(A + UCV)^{-1} = A^{-1} - A^{-1}U(C^{-1} + VA^{-1}U)^{-1}VA^{-1} \quad (22)$$

Let $\Omega_t = (M_t^{-1} + \eta P_t P_t^T)^{-1}$ and apply the Woodbury matrix identity twice, the iterative expression of $M_{t+1}$ is expressed as:

$$\begin{cases} \Omega_t = M_t - \eta M_t P_t (I + \eta P_t^T M_t P_t)^{-1} P_t^T M_t \\ M_{t+1} = \Omega_t + \eta \Omega_t Q_t (I - \eta Q_t^T \Omega_t Q_t)^{-1} Q_t^T \Omega_t \end{cases} \quad (23)$$

To ensure $M$ is a PSD matrix, we require

$$\begin{cases} \eta(P_t P_t^T - Q_t Q_t^T) + M_t^{-1} \geq 0 \\ \eta \geq 0 \end{cases} \quad (24)$$

This is a standard linear matrix inequalities (LMI). If the result of (24) is $\bar{\eta}$, then $\eta \in [0, \bar{\eta}]$ ensures that the updated $M_{t+1}$ is a PSD matrix. Thus, the LMI is solved first to find the feasible range of $\eta$.

The algorithm runs cycle by cycle. In each cycle, select $I_1$ triplet constraints, where $I_1$ equals to the number of training samples, i.e. $I_1 = N \times J$. Thus, there are $I_1$ iterations in each cycle. Then, calculate the change of the total loss function $L_k$, where $k$ is the index of cycle.

Two stop criterions are used: when $\left| \frac{L_k - L_{k-1}}{L_{k-1}} \right|$ is smaller than a predefined threshold, $\epsilon$, or when the maximum number of cycles $I_2$ is met. Note that there is a tradeoff between the execution time and accuracy when selecting $I_2$. Although the Mahalanobis matrix learning process is the most computation-intensive step, it can be executed offline.

The Mahalanobis matrix learning process can be summarized as follows:

| Algorithm 1 Mahalanobis Matrix Learning |
|---|
| 1:    Initialize Mahalanobis matrix as identity matrix |
| 2:    $k = 1, i = 1$ |
| 3:    while $k < I_2$ |
| 4:       while $i < I_1$ |
| 5:          Select the most useful triplet constraint according to [22] |
| 6:          Calculate Mahalanobis distance using (14) |
| 7:          If (15) violated |
| 8:            Then update Mahalanobis matrix using (23) |
| 9:          $i = i + 1$ |
| 10:   end |
| 11:   Calculate total loss function $L_k = \sum_i l_k(i)$ |
| 12:   If $\left| \text{abs}\left( \frac{L_k - L_{k-1}}{L_{k-1}} \right) \right|$ <threshold, break |
| 13:   $i = 1, k = k + 1$ |
| 14:  end |

### D. Template Generation

K-NN will be used to conduct the online classification. The execution time of the classification process increases rapidly as the number of instances in the training set increases. A simplification can be made if computation time requirements can no longer be met. Instead of computing the $D_M(O_i, X_i^j)$ for all members in the training set, we can select a template that represents the whole class.

First, calculate the sum of the distance, $C_j$, for a member, $X_i^{j_1}$, with all other members in the same class using

$$C_{j_1}(X_i^{j_1}) = \sum_{j=1}^{J} D_M(X_i^{j_1}, X_i^j) \quad (25)$$

Then, rank $X_i^j$ by its $C_j$ values by letting the shortest ones have the highest priorities. We can then select a predefined number out of $J$ members from the priority list to be the templates that represent the whole class. In the result section, we will compare the impact on detection accuracy when different numbers of templates are selected. The procedure of the offline training process is illustrated in Fig. 1.

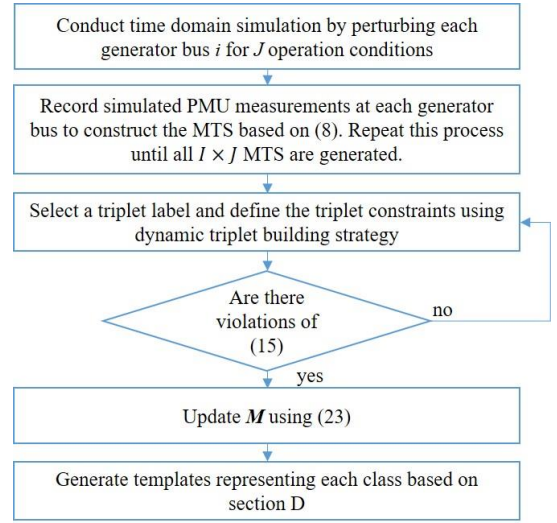

Fig. 1. Flowchart of the offline training process.

## III. MTS Online Matching

### A. Dynamic Time Warping for Out-of-sync Data

In the offline training phase, time domain simulations are used to generate the training data sets. Therefore, when selecting the detection window defined by $[t_{start} \ t_{end}]$, one can select $t_{start}$ to be right before a forced oscillation is incurred. However, in the online detection phase, it is highly likely that $t_{start}$ is not exactly the start time of the forced oscillation.

To mitigate the impact of data misalignment, we expand the training data sets by including the data of the same window length with different $t_{start}$. As shown in Fig. 2, we selected three slices of the training data using a window moving every $m$ seconds. Assuming that, in this case, a forced oscillation begins at $t = 0$. Then, the training data starts at $t = 0$, $t = m$ and $t = 2m$, respectively. The value of $m$ and the largest delay considered can be determined by engineering practice. The testing data begins at $t = d$, where $d$ is a random number simulating the detection time error. Under this situation, training data and testing data may still have different phases. Thus, the similarity of time series with different phases needs to be measured.

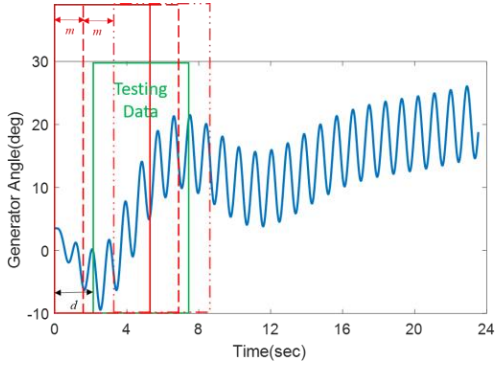

Fig. 2. Generation of training data sets.

Dynamic time warping (DTW) is an algorithm for measuring the similarity between out-of-sync time series by calculating an optimal warp path and mapping the two time series in one-to-one correspondence [24]. Traditional DTW works for univariate time series and the distance is measured by Euclidean distance. We extend DTW to MTS so that the Mahalanobis distance is applied to measure the distance between vectors instead of two points. The optimal warp path can be calculated through dynamic programming. Slightly modify the definition in (9) and (10), $x(h_1)$, $y(h_2)$ are $h_1^{th}$, $h_2^{th}$ row of two different MTS matrices $\mathbf{X}_{i_1}^{j_1}$ and $\mathbf{X}_{i_2}^{j_2}$, respectively. $d_M(x(h_1), y(h_2))$ is the Mahalanobis distance between the $h_1^{th}$ row vector of $\mathbf{X}_{i_1}^{j_1}$ and the $h_2^{th}$ row vector of $\mathbf{X}_{i_2}^{j_2}$. The calculation process of optimal warp path can be expressed as:

$$\mathbf{D}(h_1, h_2) = d_M(x(h_1), y(h_2)) + \min \begin{cases} \mathbf{D}(h_1-1, h_2-1) \\ \mathbf{D}(h_1-1, h_2) \\ \mathbf{D}(h_1, h_2-1) \end{cases}$$
$$(26)$$

where $\mathbf{D}(1,1) = d_M(x(1), y(1))$. The optimal warp path and corresponding distance can be obtained by computing all the elements in $\mathbf{D}$ matrix.

With the obtained optimal warp path $W$, the optimal alignment between the two given MTSs $\mathbf{X}_{i_1}^{j_1}$ and $\mathbf{X}_{i_2}^{j_2}$ is defined. Thus, two new MTSs with the same dimension $\widetilde{\mathbf{X}_{i_1}^{j_1}}$ and $\widetilde{\mathbf{X}_{i_2}^{j_2}}$ can be generated by

$$\begin{cases} \widetilde{\mathbf{X}_{i_1}^{j_1}}(k) = \mathbf{X}_{i_1}^{j_1}(w_1(k)) \\ \widetilde{\mathbf{X}_{i_2}^{j_2}}(k) = \mathbf{X}_{i_2}^{j_2}(w_2(k)) \end{cases} \quad (27)$$

where W is the optimal path, which shows the correspondence relationship.

$$W = \begin{pmatrix} w_1(k) \\ w_2(k) \end{pmatrix}, k = 1,2, \dots, s \quad (28)$$

$(w_1(k), w_2(k))'$ indicates the $w_1(k)^{th}$ row vector of MTS $\mathbf{X}_{i_1}^{j_1}$ corresponds to the $w_2(k)^{th}$ row vector in MTS $\mathbf{X}_{i_2}^{j_2}$.

The two new MTSs have one-to-one correspondence, thus the distance between the two original MTS $X$ and $\mathbf{X}_{i_2}^{j_2}$ can be written as

$$D(\mathbf{X}_{i_1}^{j_1}, \mathbf{X}_{i_2}^{j_2}) = \sum_{k=1}^{s} d_M(\widetilde{\mathbf{X}_{i_1}^{j_1}}(k), \widetilde{\mathbf{X}_{i_2}^{j_2}}(k)) \quad (29)$$

### B. Improved k-NN Algorithm

Using Mahalanobis distance and extended DTW, the distance of the out-of-sync MTS can be measured properly. To predict the class of the unknown objects, we first calculate $D_M(\mathbf{O}_{ClassNo}, \mathbf{X}_i^j)$, the Mahalanobis distance between the unknown object, $\mathbf{O}_{ClassNo}$, and the objects that belong to the training set, $\mathbf{X}_i^j$. In this paper, we used the $k$-NN method to assign a class membership to the unknown object by first ranking $D_M(\mathbf{O}_{ClassNo}, \mathbf{X}_i^j)$ from small to large, and then applying majority votes of the first $k$ objects. Because $k$ is a user-defined constant, in the simulation we tested the value of k on detection accuracy. For example, if $k = 1$, and the $D_M(\mathbf{O}_{ClassNo}, \mathbf{X}_{i_3}^{j_3})$ is the smallest value, then the unknown object is assigned to the class of $\mathbf{X}_{i_3}^{j_3}$, *i.e. ClassNo = 3*. This means the forced oscillation source is located on Bus 3.

As mentioned before, $k$-NN is a computational inefficiency classification approach. Typically $k$ is quite small in real application, if the length of time series is $H$, the number of instances in training set is $m_1$, the computational complexity of $k$-NN is about $O(m_1 H)$, linear in both the total number of instances and time series length. In consideration of real-time capability of the algorithm, template generation has been applied to reduce the size of training data sets. To further accelerate the execution of the algorithm, space partitioning tree is utilized [25], with which the computational complexity is reduced to $O(H \log m_1)$ at the price of more memory space.

Figure 3 summarizes the procedure of the online matching process.

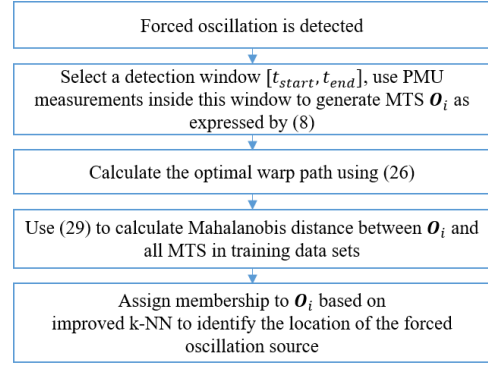

Fig. 3. Flowchart of the online matching process

## IV. NUMERICAL SIMULATION

In this section, case studies are carried out to evaluate the performance of the framework presented in Sections II and III. We perform these tests in two benchmark power systems：IEEE 39-bus system and 179-bus WECC system.

We assume that there is one PMU installed at each generator bus. PMU channels include terminal voltage (pu), active power (MW), reactive power (MVAR), absolute angle (deg) and speed (Hz) are recorded. The sampling rate of PMU is 25Hz. The simulations are conducted using TSAT and PSAT [26]. Gaussian white noises with a signal-to-noise ratio of 13dB is superimposed to the simulated PMU measurements to model the random disturbance in the PMU signal.

In this paper, we only consider cases with one forced oscillation source. For generators that are not sources of a forced oscillation, we omitted their excitation systems so the generators can be modeled as second order classical machines.

The generator serving as a forced oscillation source is modeled as a sixth order classical machine and its damping factor $D$ is set to be 0. A scaled sinusoid disturbance with the frequency of system natural mode is added to the reference signal of the excitation system to model a forced oscillation, as shown in Fig. 4.

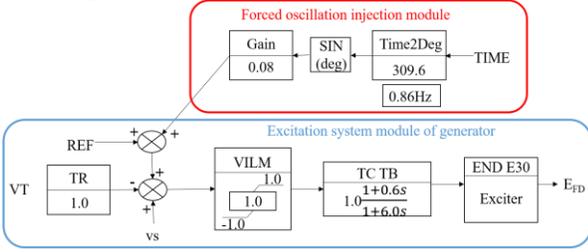

Fig. 4. Modeling of a forced oscillation

We designed 200 operation scenarios (100 for training and 100 for testing) that represent different power system loading conditions and generator parameter settings. Since the data provided by the IEEE test case is a single snapshot of the state of power system, to emulate practical situation, load fluctuations are added to all PQ loads following the Ornstein-Uhlenbeck process introduced [27], so we have

$$\dot{u}(t) = -Cu(t) + \sigma \bar{\xi} \qquad (30)$$

where $u(t)$ is the load fluctuation, $C$ is a diagonal matrix of inverse time correlations. $\bar{\xi}$ is a vector of independent standard Gaussian random variables, $\sigma = 0.01$ denotes intensity of noise.

Thus, the perturbed PQ loads are

$$S(t) = S_0(1 + u(t)) \qquad (31)$$

where $S(t) = P(t) + jQ(t)$.

### A. IEEE 39-bus System

The detailed model parameters can be found in [28]. Based on the detailed modal analysis of the system, a natural mode with the frequency at $f_0 = 1.3217Hz$ exists. Because there are 10 generators in this system, we have 10 possible forced oscillation injection points and 2000 different operation scenarios (200 operation scenarios for 10 perturbation locations) are generated.

Figure 5(a) illustrates the absolute angle measurements of Generator 1 (denoted by G1) when a forced oscillation is injected into G1 through G5, one by one. The plot shows that the dynamic response of G1 is different corresponding to different oscillation source locations. However, the oscillation tends to settle in tens of seconds, making the oscillation patterns less distinguishable over time. Thus, PMU measurements from an early phase of the oscillation are more informative for locating the oscillation source. Figure 5(b) shows the absolute angle measurements when a forced oscillation is injected into G1 under different system loading conditions. The dynamic characteristics of the oscillations are similar but their amplitudes can vary.

The forced oscillation source location problem has been converted to a multiclass classification problem with the number of classes equals to 10 for this IEEE 39-bus system. The ratio of the training data sets and testing data sets is 1:1. In (15), $\rho = 0$. Because the Mahalanobis matrix learning process is conducted offline as a non-time-critical task, we set $I_1 = 1000$ and $I_2 = 10$ for Algorithm 1.

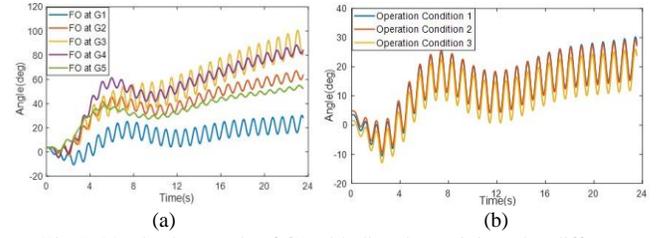

Fig.5. (a) Absolute angle of G1 with disturbance injected at different generators and (b) Absolute angle of G1 with disturbance injected at G1 for different load fluctuation

Five features for each generator are recorded. Sequential feature selection is performed to identify the most relevant three features: generator active power, absolute angle, and speed. Thus, $T = 3$, $N = 10$, and $P = N \times T = 10 \times 3 = 30$.

For training data sets, the length of the detection window is 5s, making $H = 5 \times 25$. We add the disturbance at $t = 0$s. As mentioned before, the data with a specified delay of 1 to 5 seconds are also included into the training datasets. So the total number of MTSs for training is $6 \times 100 \times 10$.

While for testing data sets, a delay $d$ is introduced to represent the uncertainty in oscillation detection. Therefore, the data fed into the classifier begins at $t = d$ seconds. In this case, we assume $d = 3.4s$ and the window size is 5s. Thus, the size of testing instance is 125×30.

Table I presents the oscillation source location performance of proposed approach. Another machine learning approach, CELL&Decision tree approach from [29], is employed as comparison. From table I, the overall accuracy of proposal is 97.8%, while the accuracy of several scenarios reaches 100%. In each scenario, the proposal outperforms the CELL&Decision method.

TABLE I PERFORMANCE TEST OF OSCILLATION SOURCE IDENTIFICATION FOR IEEE 39-BUS SYSTEM

| Generator with disturbance | Accuracy of proposal/% | Accuracy of CELL&Decision/% |
|---|---|---|
| $G_1$ | 100 | 95.7 |
| $G_2$ | 100 | 92.8 |
| $G_3$ | 100 | 98.7 |
| $G_4$ | 100 | 99.3 |
| $G_5$ | 92 | 91.3 |
| $G_6$ | 100 | 100 |
| $G_7$ | 100 | 97.3 |
| $G_8$ | 92 | 90.1 |
| $G_9$ | 94 | 93.3 |
| $G_{10}$ | 100 | 95.7 |
| Average | 97.8 | 95.4 |

### B. WECC 179-bus Test System

We choose the WECC 179-bus system with 29 generators to conduct the second test. Thus, there are 29 possible forced oscillation source locations (i.e., $N = 29$). The test case library provided by the IEEE PES Task Force on Oscillation Source Location [30] is used as the baseline setting.

Because the eigenvalue analysis of the WECC 179-bus system shows that an oscillation mode exists at the frequency of 0.86 Hz, a scaled 0.86 Hz sinusoid disturbance is added to the reference signal of the excitation system to model the forced oscillation. In total, 5800 samples are generated to cover 200 operation scenarios at each of the 29 possible perturbation locations. Let each PMU has 3 most informative channels, we have $P = N \times T = 29 \times 3 = 87$. Let the length of the detection window be 5s, $H = 5 \times 25$. Assume that the disturbance happens at $t = 0\,s$, we construct $6 \times 100 \times 29$ training data sets by considering a detection delay ranging from 0 to 5 seconds.

First, we investigate the impact of the number of nearest neighbors on oscillation detection error. The Metric learning process is conducted to train Mahalanobis matrix using the whole training data sets. Five values of oscillation detection error from 0 to 5 seconds are selected where the case with $d = 0\,s$ is considered to produce the ideal result. The improved $k$-NN classification is then used to assign a category label to each testing instance, in which the number of the nearest neighbors $k$ changes from 1 to 10.

After running the simulation using the whole testing data sets for each combination of $d$ and $k$, we obtain the precision of classification with respect to $k$ for the 6 detection errors, as shown in Fig. 6. We have the following observations:

- For the ideal case ($d = 0s$, $k$=1), among all 29 scenarios, the highest accuracy is 100%, while the lowest accuracy is 93%.
- The online matching time is 28.015s for each testing instance on a PC (3.7Ghz Intel Core i7 processer with 32GB RAM).
- For all cases, the location accuracy is above 95%. The highest accuracy is 97.91% when $k = 1$, $d = 0s$ while the lowest accuracy is 95.14% when $k = 10$, $d = 2.8s$.
- For a given $d$, there is no significant diffidence in precision of classification when $k$ changes because when $k$ varies from 1 to 10, the precision variation at each $d$ is within 2.27%.
- When $k$ increases, the precision of classification decreases. This is because $k$ is larger, the instances far from the actual will put more influence on the classifier, leading to a lower classification accuracy.

From the observation, for simplicity, $k$ is set as 1 when conducting online matching in the rest of the case studies.

As shown in Fig. 5(a), the measurements in an earlier phase of an oscillation pose more distinct information about the oscillation source location than those from later phases. However, it is not practical to count on the very beginning of an oscillation being captured by the sensors. When comparing the six curves in Fig. 6, we see that when $d = 0s$, the classification achieves the highest accuracy, except two cases (e.g., $k = 8, k = 10$). As can be seen from fig.6, delay $d$ increases, the location accuracy declines. This is because the proposed approach localizes oscillation sources mainly based on the dynamic characteristics of features in the transient process. When the oscillation tends to be stable, the characteristic of features becomes indistinct. So, the more accurate of the oscillation is detected, the higher is the location accuracy.

Because forced oscillation can attenuate rapidly once the disturbance source is removed, the sooner the oscillation source is located and removed, the better. On the other hand, $k$-NN is a classification approach in which efficiency of classification rapidly decreases as the number of instances in training data set increases. Thus, we propose the template generation and classification method to increase the algorithm efficiency and shorten the computation time. Two criteria are used to evaluate the performance of the template classification method: the classification accuracy and the online matching time. The ideal case, where all the instances in the training data sets are used, is set as the benchmark case.

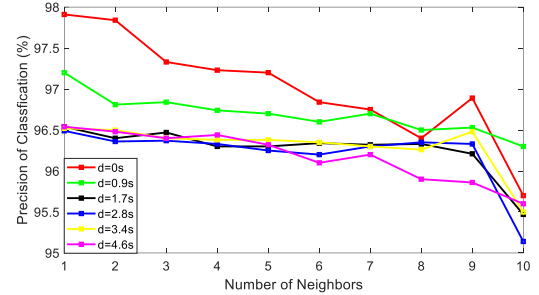

Fig.6. Precision of classification for different detection errors

Templates have been generated based on section II part D for each class, and $k$-NN with $k$=1 is applied. It means a testing instance $O_i$(whose class label is unknown) shall be assigned to a class represented by template $X_{i_T}^{j_T}$ if distance of $O_i$ with $X_{i_T}^{j_T}$ is minimum compared with the ones between $O_i$ and all other templates. The simulation results are listed in table II.:

TABLE II THE PERFORMANCE OF DIFFERENT NUMBER OF TEMPLATES

|  | Precision of Classification(%) | Time for Online Matching(s) |
|---|---|---|
| Benchmark | 97.7 | 28.015 |
| 1 template | 95.94 | 0.29 |
| 2 templates | 95.94 | 0.565 |
| 3 templates | 95.52 | 0.875 |
| 4 templates | 96.05 | 1.085 |
| 5 templates | 96.21 | 1.305 |
| 6 templates | 95.75 | 1.655 |

Various number of templates are considered and tested. As can be seen from the table, the precision of classification keeps high even with much lesser instances in training data sets. In the meantime, the time needed for online matching declines dramatically. When using 1 to 6 templates for each scenario for online matching, the time needed ranges from 0.29s to 1.655s, approximately 95% less than that of the benchmark takes.

Therefore, by implementing template classification, the algorithm can locate the oscillation source in an online fashion with a relatively high accuracy while satisfying the timeliness requirement. As analyzed in section III part B, the computational complexity is $O(Hlogm_1)$ with $H$ as the length of time series and $m_1$ as the number of instances in training set. For a larger system with 500 generators, the online matching time using one template is approximately 5 seconds, showing that it is applicable for large system applications. Of course, using more instances in the training data sets leads to a higher classification accuracy so there is a tradeoff between required classification precision and targeted execution time when selecting the number of templates.

### C. Analysis of mis-classified cases

In this section, we investigate the mis-classified cases. Since FO sources geographically close to each other can excite oscillations bearing similar signatures, the identified source may locate at a bus that is close to the actual. In the WECC-179 bus system, when $d = 1.7s$, $k = 1$, and the FO source located at Generator 9 (highlighted with a red pentagram) in the 100 testing instances, three misclassified cases are found, in which Generator 4 (highlighted by an orange circle) is identified as the source instead of 9. As shown in Fig. 7 (a), the two generators are geographically very close to each other. Classification confusion matrix for 29 scenarios are plotted in Fig.7 (b), in which the x axis and y axis are the generator bus number. Similar misclassification cases can also be found in cases for other FO locations. Therefore, we conclude that even for the missing cases, the search range can be narrowed down to the neighboring buses of the identified one.

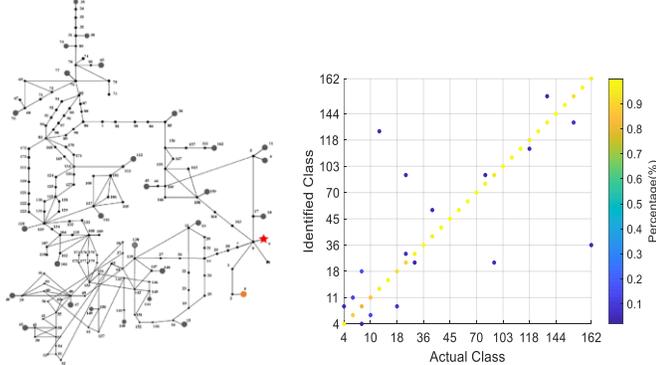

Fig.7. WECC 179-bus power system: (a) complete topology [30]; (b)Classification confusion matrix

## V. CONCLUSION AND FUTURE WORK

In this paper, we proposed a machine-learning based method for locating FO sources. Simulation results in IEEE 39-bus system and WECC 179-bus system demonstrated that, by using a combination of metric learning, DTW, and the improved k-NN algorithms, the proposed method achieved reduced computational time (within a few seconds), improved classification precision (95%+), and robustness when handling data out-of-sync issues (up to 5 second). In our subsequent work, we will focus on the detection and identification of forced oscillation and poorly damped oscillation as well as the remedial actions for mitigating their impacts on power grid operation.